\documentclass[prd,twocolumn,showpacs,floatfix,amsmath,amssymb,floatfix]{revtex4}
\usepackage{graphicx,color,dcolumn,booktabs,bm}
\usepackage{longtable,lscape}
\usepackage{txfonts}
\usepackage{overpic}
\usepackage{amssymb}
\usepackage{indentfirst}
\usepackage{feynmf}   
\usepackage{slashed}  
\usepackage{cases}
\usepackage{color}
\usepackage{multirow}
\usepackage{graphicx,color,dcolumn,booktabs,bm}
\usepackage{epsfig,dsfont,amssymb,amsmath,amsfonts,amsbsy,mathrsfs}

\graphicspath{{Figures/}} %

\begin{document}

\title{The hidden symmetries in the PMNS matrix and the light sterile neutrino(s)}

\vspace{1cm}

\author{ Hong-Wei Ke$^{1}$\footnote{khw020056@hotmail.com}, Jia-Hui Zhou$^{1}$, Shuai
Chen$^{1}$, Tan Liu$^{1}$ and Xue-Qian
Li$^2$\footnote{lixq@nankai.edu.cn}}

\affiliation{  $^{1}$ School of Science, Tianjin University, Tianjin 300072, China \\
  $^{2}$ School of Physics, Nankai University, Tianjin 300071, China }

\vspace{12cm}

\begin{abstract}
The approximately symmetric form of the PMNS matrix suggests that
there could exist a hidden symmetry which makes the PMNS matrix
different from the CKM matrix for quarks. In literature, all the
proposed fully symmetric textures exhibit an explicit $\mu-\tau$
symmetry in addition to other symmetries which may be different
for various textures. Observing obvious deviations of the
practical PMNS matrix elements from those in the symmetric
textures, there must be a mechanism to distort the symmetry. It
might be due to existence of light sterile neutrinos. As an
example we study the case of the Tribimaximal texture and propose
that its apparent symmetry disappears due to existence of a
sterile neutrino.



\end{abstract}

\maketitle

Numerous experiments which are carried out in past several decades
make the behaviors of neutrinos understandable. As is commonly
accepted, mixing among different flavors of leptons are due to the
mismatch between the mass eigenstates and flavor eigenstates, it
is the same as the quark case, but different in structures. To
bring the weak interaction eigenstates (flavor) to the physical
ones (mass), the Pontecorvo-Maki-Nakawaga-Sakata (PMNS)
matrix\cite{Pontecorvo:1967fh,Maki:1962mu} should be introduced.
If there are only three active neutrinos the mixing matrix is
written as
    \begin{equation}\label{M1}
      V=\left(\begin{array}{ccc}
        V_{11} &V_{12} &V_{13} \\
         V_{21} &   V_{22} &  V_{23}\\
          V_{31} & V_{32} & V_{33}
      \end{array}\right).
  \end{equation}
Generally there are four independent parameters, namely three
mixing angles and one CP-phase. There are various schemes to
parameterize the matrix  in literature. For example, the Chau-Keung(CK)
parametrization \cite{Chau:1984fp} is
 \begin{equation}\label{M2}
      V=\left(\begin{array}{ccc}
        c_{12}c_{13} & s_{12}c_{13} & s_{13}\\
       -c_{12}s_{23}s_{13}e^{i \delta} -s_{12}c_{23} & -s_{12}s_{23}s_{13}e^{i\delta}+c_{12}c_{23} & s_{23}c_{13}\\
      -c_{12}s_{23}s_{13}e^{i\delta} + s_{12}s_{23} & -s_{12}s_{23}s_{13}e^{i\delta} -c_{12}s_{23}& c_{23}c_{13}
      \end{array}\right),
\end{equation}
where $s_{jk}$ and $c_{jk}$ denote $\sin\theta_{jk}$ and
$\cos\theta_{jk}$ with $j,k=1,2,3$.

The measured values of the PMNS matrix exhibit an approximately symmetric form
which may hint that the practical matrix originates from a high symmetry, but is distorted by
some mechanisms. Indeed, one of the physics achievements of the 20th century convinces us that symmetry and symmetry
breaking compose the main picture of the nature, so one may reasonably expect that an underlying
symmetry determines the mixing matrix of leptons which later is distorted somehow. Lam has
shown this possibility in terms of the group theory\cite{Lam:2011ip} where the CKM and PMNS matrices are separately
resulted via different routes to break the large symmetry. In Lam's scheme, the resultant PMNS still possesses an
obvious symmetry with $\theta_{13}$ strictly being zero, therefore to reach the practical PMNS a further symmetry breaking
is needed. It is natural to ask if one can provide a reasonable mechanism to explain the distortion.

Meanwhile some phenomenological symmetries are observed, such
as the quark-lepton complementarity and
self-complementarity\cite{Minakata:2004xt,Raidal:2004iw,Altarelli:2009kr,Zheng:2011uz,Zhang:2012pv,Zhang:2012zh,Haba:2012qx},
$\mu-\tau$ symmetry\cite{Wolfenstein:1978uw}. But all those symmetries are only approximate, so it also implies that there
should be some  mechanisms to result in
their deviations from exact symmetric forms.

The high-precision measurements
\cite{Eguchi:2002dm,Ahmed:2003kj,Ahn:2002up,Jung:2001dh}
determines $\theta_{12}\approx34^\circ$ and
$\theta_{23}\approx45^\circ$ and a small, but non-zero
$\theta_{13}$. The values could be traced to a mixing pattern with
high symmetries i.e. for example the tribimaximal (TB) mixing
pattern\cite{Harrison:1999cf,Xing:2002sw,Fritzsch:1995dj,Wolfenstein:1978uw}
which is one of the possible symmetric textures
\begin{equation}\label{M3}
      V_{TB}=\left(\begin{array}{ccc}
        \sqrt{\frac{2}{3}} &\frac{1}{\sqrt{3}} &0 \\
         -\frac{1}{\sqrt{6}} &  \frac{1}{\sqrt{3}} &  \frac{1}{\sqrt{2}}\\
          \frac{1}{\sqrt{6}} & -\frac{1}{\sqrt{3}} & \frac{1}{\sqrt{2}}
      \end{array}\right).
  \end{equation}
which means that $\theta_{12}=35.26^\circ$,
$\theta_{23}=45^\circ$ and $\theta_{13}=0^\circ$ in the adopted parametrization. In this
scenario the $\mu-\tau$ symmetry holds and in the
mass eigenstate $\nu_2$,  $\nu_e$, $\nu_\mu$
and $\nu_\tau$ have the same probability.

However the measurements of the accelerator and reactor
neutrino oscillation
experiments\cite{Abe:2011fz,Abe:2011sj,Adamson:2011qu,Ahn:2012nd,An:2012eh}
determine $\theta_{12}=(33.65^{+1.11}_{-1.00})^\circ$,
$\theta_{23}=(38.41^{+1.40}_{-1.21})^\circ$ and
$\theta_{13}=(8.93^{+0.46}_{-0.48})^\circ$. It is noted that the
values are set based on the scenario for three generations of neutrinos. Apparently the
TB mixing patterns decline from the data. One may ask whether the symmetries
in the TB mixing patterns should be abandoned? Even though it is too early to
make a definite conclusion yet, there exists a possibility that those symmetries still hold, whereas the matrix might be distorted
from the symmetric form by new physics.

Recently, the anomalies of short-baseline  neutrino
experiments\cite{Sorel:2003hf,Maltoni:2007zf,Karagiorgi:2009nb}
hint that there may exist light sterile neutrinos which  mix with
the active ones. Moreover Schechter and Valle
\cite{Schechter:1980gr,Schechter:1981cv} also discussed the (n,m)
model which suggested importance of getting sterile neutrino
involved. If this picture indeed works, the existence of sterile
neutrinos would play a role to make the mixing matrix being in the
superficial form where the original symmetries are just hidden
somehow or slightly broken. In our earlier work \cite{Ke:2014hxa},
we proposed that the quark-lepton complementarity and
self-complementarity\cite{Minakata:2004xt,Raidal:2004iw,Altarelli:2009kr,Zheng:2011uz,Zhang:2012pv,Zhang:2012zh,Haba:2012qx}
still hold  and involvement of the sterile neutrino(s) distorts
them to be approximate. Along the same line, one may ask if the
symmetries in the TB mixing patterns are distorted by existence of
sterile neutrinos. In other words, when PMNS is extended to a
$n\times n$ matrix, the symmetries in TB mixing patterns are exact
or just slightly broken, but in the left-upper  $3\times 3$ block
of the generalized PMNS matrix which corresponds to the
experimentally observed values, the symmetries are no longer
exhibited at all. In an explicit statement, the symmetries which
are shown in the original $3\times 3$ texture, still hold or
slightly broken, but as mixing between the sterile and active
neutrinos exists, the apparent symmetric form is lost.

In this letter we use this picture to study the neutrino mixing matrix i.e.
supposing that the symmetries in the
tribimaximal (TB) mixing patterns are exact or nearly exact, but
a mixing among active neutrinos and sterile neutrinos
causes the difference between the TB mixing matrix and data.

Here, we only consider the simplest scheme where three active neutrinos plus
one sterile neutrino  are involved, namely the (3+1) scheme and show that it indeed works.


As a (or a few) light sterile neutrino joins the game, the neutrino mass
matrix would turn into a $4\times 4$ form. If charged-lepton mass
matrix is still diagonal \cite{Liu:2013oxa}, a real neutrino
mass matrix which may possess certain symmetries and manifest in special textures as suggested by
some authors, is written as
\begin{equation}\label{M4}
     V_{4\times 4}=\left(\begin{array}{cccc}
        A& B& C&D\\
     B & E &F&G\\
    C &F & H&
    I\\
      D&G&I&J
      \end{array}\right),
\end{equation}
can be diagonalized by  a $4\times 4$ PMNS-like unitary
matrix\cite{Girardi:2014wea,Kisslinger:2013sba}
\begin{widetext}
\begin{equation}\label{M5}
     V_{4\times 4}=\left(\begin{array}{cccc}
        c_{12}c_{13}& s_{12}c_{13} & s_{13} &0\\
     -s_{12}c_{23}-c_{12}c_{23}s_{13} & c_{12}c_{23}-s_{12}s_{23}s_{13}  & c_{13} s_{23}&0\\
      s_{12}s_{23}- c_{12}c_{23}s_{13} &  -c_{12}s_{23}-s_{12}c_{23}s_{13}&
      c_{23}c_{13} &0\\
      0&0&0&1
      \end{array}\right)
      \left(\begin{array}{cccc}
       \mathrm{cos}{\alpha} &0 &0&\mathrm{sin}{\alpha} \\
        0 & 1 &  0& 0 \\
        0 & 0 &1& 0\\
         -\mathrm{sin}{\alpha}& 0& 0&\mathrm{cos}{\alpha}
      \end{array}\right)\\\\
      \left(\begin{array}{cccc}
      1 &0 &0&0 \\
        0 &\mathrm{cos}{\beta} &  0&\mathrm{sin}{\beta} \\
       0 & 0 &1& 0\\
          0& -\mathrm{sin}{\beta}& 0&\mathrm{cos}{\beta}
      \end{array}\right)\left(\begin{array}{cccc}
       1  &0&0&0 \\
      0 &  1 &  0& 0 \\
      0 & 0 &\mathrm{cos}{\gamma}&\mathrm{sin}{\gamma}\\
          0& 0&-\mathrm{sin}{\gamma}&\mathrm{cos}{\gamma}
      \end{array}\right).
\end{equation}
\end{widetext}
For convenience of later discussions we still set $\theta_{13}=0^\circ$ and keep
$\theta_{12}$ and $\theta_{23}$ in the expression. In the
expression the first matrix is the original PMNS matrix and the
successive three matrices correspond to the mixing of the sterile
neutrino with the three active flavors respectively. For this case
one can find
$|V_{13}|=|s_{\gamma}(c_{\beta}s_{\alpha}c_{12}+s_{\beta}s_{12})|$.
We first constrain ourselves to the simplest setting
$\alpha=\beta=\gamma$. By the same procedures as done in
Ref.\cite{Ke:2014hxa} we fix $\alpha=\beta=\gamma=18.001^\circ$
and now the $4\times 4$ matrix reads
\begin{equation}\label{M6}
      |V_{4\times 4}|=\left(\begin{array}{cccc}
       \mathbf{ 0.776531}& \mathbf{0.471115} &\mathbf{ 0.12969}&0.397912\\
    \mathbf{0.388265}& \mathbf{0.588078} & \mathbf{0.654438}&0.274091\\
    \mathbf{ 0.388265}& \mathbf{0.588078} & \mathbf{0.690553}&0.162945\\
      0.309031&0.293904&0.279518&0.860226
      \end{array}\right).
      \end{equation}

Comparing with the experimentally determined $3\times 3$ PMNS matrix
 $V_{PMNS}$ Ref.\cite{Zhang:2012pv}
    \begin{equation}\label{M7}
      |V_{PMNS}|=\left(\begin{array}{ccc}
       0.822^{+0.010}_{-0.011} &0.547^{+0.016}_{-0.015} & 0.155^{+0.008}_{-0.008}\\
       0.451^{+0.014}_{-0.014} & 0.648^{+0.012}_{-0.014}&0.614^{+0.019}_{-0.017}\\
      0.347^{+0.016}_{-0.014} &0.529^{+0.015}_{-0.014}
        &0.774^{+0.013}_{-0.015}
      \end{array}\right),
  \end{equation}
the left-upper $3\times 3$ block of the $V_{4\times 4}$ matrix  is
nearly consistent with that form and $|V_{13}|$ is no longer zero.
By fitting data, we only fit matrix elements $V_{11}$, $V_{12}$,
$V_{13}$, $V_{23}$ and $V_{33}$ which are free of CP phase.
Because so far the CP phase has not been experimentally determined
and no even any hint is available, we cannot expect to extract
information on $\delta$ from data. In our previous work, by the
$\chi^2$ analysis, we find that the CP phase  $\delta$ is within a
region close to zero, but all of the theoretical predictions are
waiting for probes of future more precise experiments. Therefore
in this work, the derived mixing matrix is real.

The obtained matrix elements are close to the data, but still not
fully satisfactory yet. If slightly breaking the $\mu-\tau$
symmetry, the situation would be further improved. Concretely, we introduce a small variation $\epsilon$ to
$\theta_{23}$, i.e. let the original $\theta_{23}$ slightly
deviate from $45^\circ$. Obviously, $\epsilon$ should be
determined by fitting data. Thus, one further fix the values
as
      $\theta_{23}=39.3999^\circ$ and
      $\alpha=\beta=\gamma=17.777^\circ$,
      \begin{equation}\label{M8}
     |V_{4\times 4}|=\left(\begin{array}{cccc}
       \mathbf{ 0.777551}&\mathbf{ 0.473604} & \mathbf{0.126298}&0.393906\\
    \mathbf{0.424789}& \mathbf{0.642425} & \mathbf{0.585201}&0.25374\\
    \mathbf{ 0.348925}& \mathbf{0.527692} & \mathbf{0.751624}&0.186692\\
      0.305319&0.29074&0.276858&0.863481
      \end{array}\right).
      \end{equation}
It is noted that using the mechanism of involving a sterile
neutrino and a slight $\mu-\tau$ symmetry breaking, the data can
be well explained. The scheme is equivalent to refitting the data
with two independent parameters instead of one as done above. It
is noted that in this scheme the transformation matrix remains
unitary. It is also natural to expect that the symmetry breaking degree
could be at order of ${\cal O}({m_{\mu}\over m_{\tau}})$,
concretely, as we refit the data to obtain $\epsilon$, it is
$\epsilon\sim(\sin 45^{\circ}- \sin 39.3999^{\circ})/\sin 45^{\circ}\approx
2(m_{\mu}/m_{\tau})$, which indeed is at the expected order.

In fact as the $\mu-\tau$ symmetry is lifted, the constraint
$\alpha=\beta=\gamma$ is no longer valid and one will obtain a new
$|V_{4\times 4}|$
 \begin{equation}\label{M9}
     |V_{4\times 4}|=\left(\begin{array}{cccc}
       \mathbf{ 0.792861}&\mathbf{ 0.501251} & \mathbf{0.134394}&0.319482\\
    \mathbf{0.431412}& \mathbf{0.633394} & \mathbf{0.561101}&0.312829\\
    \mathbf{ 0.357905}& \mathbf{0.525473} & \mathbf{0.732239}&0.244147\\
      0.239079&0.267277&0.36186&0.860501
      \end{array}\right),
      \end{equation}
in the new setting, we have  $\theta_{23}=39.680^\circ$,
      $\alpha=13.823^\circ, \beta=15.977^\circ$, and
      $\gamma=22.808^\circ$.

Alternatively one may choose another scheme, that he keeps
$\theta_{23}=45^\circ$, $\theta_{13}=0^\circ$ and lets the original
$\theta_{12}$ slightly deviate from $35.26^\circ$. In that scheme, the best
fitting values are $\alpha=5.126^\circ, \beta=27.748^\circ,
\gamma=23.220^\circ$ and $\theta_{12}=37.370^\circ$. Namely,
the value of $\alpha$ is smaller than that obtained in other cases whereas
$\beta$ and $\gamma$ are larger. The corresponding $|V_{4\times
4}|$ matrix is
 \begin{equation}\label{M10}
     |V_{4\times 4}|=\left(\begin{array}{cccc}
 \mathbf{ 0.791554}  &  \mathbf{0.504097 } &  \mathbf{0.136193} &  0.31746\\
\mathbf{ 0.427469}  & \mathbf{0.51519} & \mathbf{0.560057 } &  0.488043\\
\mathbf{0.427469 } & \mathbf{0.51519}  &  \mathbf{0.739606}  &  0.0695213\\
 0.0893532 & 0.463729 &  0.347522 &  0.810062
\end{array}\right),
      \end{equation}
where the matrix elements $V_{21}$ and $V_{22}$ are still equal to
$V_{31}$ and $V_{32}$ respectively.

In Ref.\cite{Mention:2011rk}, the reactor antineutrino anomaly is
studied where $|\Delta m_{41}|^2>1.5$eV$^2$ and
$\sin^2(2\theta_{14})=0.14\pm 0.08$ are obtained, the resultant
mixing element $\sin^2(2\theta_{14})$  is a bit lower than our
estimation $0.22\sim 0.34$. Our prediction on $V_{e4}^2=0.102 \sim
0.155$  is also $3\sim 5$ times larger than  $0.024\sim
0.033$\cite{Giunti:2013aea} determined by the data of
short-baseline neutrino oscillation, but consistent with the
number of 0.15\cite{Giunti:2011cp} obtained by the data of Gallium
radioactive source experiments and 0.13\cite{Giunti:2011cp} gained
by fitting the data of the measurement on $\nu_e+^{12}C\rightarrow
^{12}N_{g.s.}+e^-$. In Ref.\cite{Giunti:2011cp} the authors
indicate that the large $V^2_{e4}$ in the Neutrinoless
$\beta\beta$-decay is favored to guarantee a low value of $|\Delta
m^2_{41}|$ which are required by the cosmological constraints
\cite{Hamann:2010bk,Giusarma:2011ex}.

It is noted that the mixing angles between the active neutrinos
and sterile neutrino are relatively large, so that in the
resultant upper-left $3\times 3$ block in Eq.(\ref{M6}),
Eq.(\ref{M8}), Eq.(\ref{M9}) or Eq.(\ref{M10}) the unitarity is
violated to some extent. Generally, introducing a sterile neutrino
which mixes with the active ones, the new $4\times 4$ matrix is
unitary and the upper $3\times 3$ sub-matrix is no longer unitary.
But the data seem to favor an almost unitary $3\times 3$ matrix
which corresponds to the direct measurements. Comparing the
results given in Refs.\cite{Antusch:2014woa,Escrihuela:2015wra}
where  the authors studied the violation of the unitarity of the
neutrino mixing matrix, our results manifest a larger violation
beyond  the constrains  set in \cite{Antusch:2014woa}. Since the
errors of concerned experiments which measure properties of
neutrinos are still large,  our results, so far, do not
drastically conflict with data, i.e. are still tolerable. However,
if the future experiments further confirm the constraints with
higher accuracy, our scheme that one can distort the tribimaximal
mixing pattern by introducing just one sterile neutrino should be
abandoned. In fact, there are lots of alternations to remedy the
scheme, for example, one may invoke a complex 3 (active
neutrinos)+n (sterile neutrinos) mixing scheme and/or choose other
symmetrical  PMNS textures, the mixing angles between active
neutrinos sterile neutrinos  could be very different, namely
violation of unitarity of the $3\times 3$ matrix  may be
alleviated and meanwhile the data can still be realized.

{As a brief conclusion, in this work we do not intend to make a
complete analysis based on the group theory as done by many
authors\cite{Babu:2002dz,Morisi:2012fg,Morisi:2013qna}, instead we
have carried out a phenomenological study. Recently, many authors
suggest that existence of light sterile neutrinos may lead to
solutions for some phenomenological problems where the theoretical
predictions deviate from experimental data. As indicated in the
introduction, the practical PMNS matrix is obviously distorted
from the symmetrical form, therefore it is natural to conjecture
if introducing a sterile neutrino can make up the gap between the
proposed symmetric texture and the practical data.}

{Here as an example we discuss the case of the TB mixing patterns.
First, as only one light sterile neutrino is included, the matrix
$V_{TB}$ is extended to a $V_{4\times 4}$. The mixing angles
between the sterile neutrino with the active ones in the $4\times
4$ extended PMNS matrix $V_{PMNS}$ are fixed by fitting data. The
results show that the element $V_{13}$ is non-zero and close to
the newly measured value. Meanwhile the unitarity of the $3\times
3$ is violated to an uncomfortable degree, but not too much beyond
the tolerable range yet. It is noted that in this work, we take
the tribimaxial texture as the subject, and employ the simplest
3+1 model to carry out the computation. We expect the future
experiments to verify or negate the simplest scheme, and then we
may turn to invoke more complicated cases, namely, if a more
complicated scheme and/or another symmetrical PMNS texture (not
the TB) are taken the results might be closer to the real data and
the unitarity of the $3\times 3$ matrix should be approximately
retained.}

\section*{Acknowledgement}
 This work
is supported by the National Natural Science Foundation of China
(NNSFC) under the contract No. 11375128 and 11135009.

\appendix



\begin{thebibliography}{99}
\bibitem{Pontecorvo:1967fh}
  B.~Pontecorvo,
  Sov.\ Phys.\ JETP {\bf 26}, 984 (1968)  [Zh.\ Eksp.\ Teor.\ Fiz.\  {\bf 53}, 1717 (1967)].

\bibitem{Maki:1962mu}
  Z.~Maki, M.~Nakagawa and S.~Sakata,
  Prog.\ Theor.\ Phys.\  {\bf 28}, 870 (1962).  
\bibitem{Chau:1984fp}
  L.~L.~Chau and W.~Y.~Keung,
   Phys.\ Rev.\ Lett.\  {\bf 53}, 1802 (1984).  

\bibitem{Lam:2011ip}
 C.~S.~Lam,
   arXiv:1105.4622 [hep-ph]; 
  C.~S.~Lam,
  Phys.\ Rev.\ D {\bf 83}, 113002 (2011)  [arXiv:1104.0055 [hep-ph]].  

\bibitem{Zhang:2012pv}
  Y.~Zhang, X.~Zhang and B.~-Q.~Ma,
   Phys.\ Rev.\ D {\bf 86}, 093019 (2012)  [arXiv:1211.3198 [hep-ph]].  



\bibitem{Minakata:2004xt}
  H.~Minakata and A.~Y.~.Smirnov,
  Phys.\ Rev.\ D {\bf 70}, 073009 (2004)  [hep-ph/0405088].  


\bibitem{Raidal:2004iw}
  M.~Raidal,
  Phys.\ Rev.\ Lett.\  {\bf 93}, 161801 (2004)  [hep-ph/0404046].  

\bibitem{Altarelli:2009kr}
  G.~Altarelli, F.~Feruglio and L.~Merlo,
  JHEP {\bf 0905}, 020 (2009)  [arXiv:0903.1940 [hep-ph]];  
   G.~Altarelli and D.~Meloni,
  J.\ Phys.\ G {\bf 36}, 085005 (2009)  [arXiv:0905.0620 [hep-ph]];  
  R.~de Adelhart Toorop, F.~Bazzocchi and L.~Merlo,
  JHEP {\bf 1008}, 001 (2010)  [arXiv:1003.4502 [hep-ph]];  
  G.~Altarelli, F.~Feruglio, L.~Merlo and E.~Stamou,
  JHEP {\bf 1208}, 021 (2012)  [arXiv:1205.4670 [hep-ph]].  



\bibitem{Zheng:2011uz}
  Y.~-j.~Zheng and B.~-Q.~Ma,
  Eur.\ Phys.\ J.\ Plus {\bf 127}, 7 (2012)  [arXiv:1106.4040 [hep-ph]];  
  X.~Zhang and B.~-Q.~Ma,
    Phys.\ Rev.\ D {\bf 86}, 093002 (2012)  [arXiv:1206.0519 [hep-ph]];  
  H.~Qu and B.~-Q.~Ma,
   Phys.\ Rev.\ D {\bf 88}, 037301 (2013)  [arXiv:1305.4916 [hep-ph]].  



\bibitem{Zhang:2012zh}
  X.~Zhang, Y.~-j.~Zheng and B.~-Q.~Ma,
  Phys.\ Rev.\ D {\bf 85}, 097301 (2012)
  [arXiv:1203.1563 [hep-ph]].



\bibitem{Haba:2012qx}
  N.~Haba, K.~Kaneta and R.~Takahashi,
  Europhys.\ Lett.\  {\bf 101}, 11001 (2013)  [arXiv:1209.1522 [hep-ph]].  

\bibitem{Harrison:1999cf}
  P.~F.~Harrison, D.~H.~Perkins and W.~G.~Scott,
  Phys.\ Lett.\ B {\bf 458}, 79 (1999)
  [hep-ph/9904297];
  P.~F.~Harrison, D.~H.~Perkins and W.~G.~Scott,
  Phys.\ Lett.\ B {\bf 530}, 167 (2002)
  [hep-ph/0202074].

\bibitem{Xing:2002sw}
  Z.~z.~Xing,
  Phys.\ Lett.\ B {\bf 533}, 85 (2002)
  [hep-ph/0204049].

\bibitem{Fritzsch:1995dj}
  H.~Fritzsch and Z.~Z.~Xing,
  Phys.\ Lett.\ B {\bf 372}, 265 (1996)
  [hep-ph/9509389];
  H.~Fritzsch and Z.~z.~Xing,
  Phys.\ Lett.\ B {\bf 440}, 313 (1998)
  [hep-ph/9808272];
  H.~Fritzsch and Z.~z.~Xing,
  Prog.\ Part.\ Nucl.\ Phys.\  {\bf 45}, 1 (2000)
  [hep-ph/9912358].


\bibitem{Wolfenstein:1978uw}
  L.~Wolfenstein,
  Phys.\ Rev.\ D {\bf 18}, 958 (1978).  



\bibitem{Eguchi:2002dm}
  K.~Eguchi {\it et al.}  [KamLAND Collaboration],
  Phys.\ Rev.\ Lett.\  {\bf 90}, 021802 (2003)  [hep-ex/0212021].  

\bibitem{Ahmed:2003kj}
  S.~N.~Ahmed {\it et al.}  [SNO Collaboration],
  Phys.\ Rev.\ Lett.\  {\bf 92}, 181301 (2004)  [nucl-ex/0309004].  

\bibitem{Ahn:2002up}
  M.~H.~Ahn {\it et al.}  [K2K Collaboration],
  Phys.\ Rev.\ Lett.\  {\bf 90}, 041801 (2003)  [hep-ex/0212007].  

\bibitem{Jung:2001dh}
  C.~K.~Jung, C.~McGrew, T.~Kajita and T.~Mann,
  Ann.\ Rev.\ Nucl.\ Part.\ Sci.\  {\bf 51}, 451 (2001).  



\bibitem{Abe:2011sj}
  K.~Abe {\it et al.}  [T2K Collaboration],
  Phys.\ Rev.\ Lett.\  {\bf 107}, 041801 (2011)  [arXiv:1106.2822 [hep-ex]].  
\bibitem{Adamson:2011qu}
  P.~Adamson {\it et al.}  [MINOS Collaboration],
   Phys.\ Rev.\ Lett.\  {\bf 107}, 181802 (2011)  [arXiv:1108.0015 [hep-ex]].  
\bibitem{Abe:2011fz}
  Y.~Abe {\it et al.}  [DOUBLE-CHOOZ Collaboration],
   Phys.\ Rev.\ Lett.\  {\bf 108}, 131801 (2012)  [arXiv:1112.6353 [hep-ex]].  

\bibitem{An:2012eh}
  F.~P.~An {\it et al.}  [DAYA-BAY Collaboration],
  Phys.\ Rev.\ Lett.\  {\bf 108}, 171803 (2012)  [arXiv:1203.1669 [hep-ex]].  

\bibitem{Ahn:2012nd}
  J.~K.~Ahn {\it et al.}  [RENO Collaboration],
   Phys.\ Rev.\ Lett.\  {\bf 108}, 191802 (2012)  [arXiv:1204.0626 [hep-ex]].  



\bibitem{Sorel:2003hf}
  M.~Sorel, J.~M.~Conrad and M.~Shaevitz,
  Phys.\ Rev.\ D {\bf 70}, 073004 (2004)
  [hep-ph/0305255].

\bibitem{Maltoni:2007zf}
  M.~Maltoni and T.~Schwetz,
  Phys.\ Rev.\ D {\bf 76}, 093005 (2007)
  [arXiv:0705.0107 [hep-ph]].

\bibitem{Karagiorgi:2009nb}
  G.~Karagiorgi, Z.~Djurcic, J.~M.~Conrad, M.~H.~Shaevitz and M.~Sorel,
  Phys.\ Rev.\ D {\bf 80}, 073001 (2009)
  [Erratum-ibid.\ D {\bf 81}, 039902 (2010)]
  [arXiv:0906.1997 [hep-ph]].

\bibitem{Schechter:1980gr}
  J.~Schechter and J.~W.~F.~Valle,
  Phys.\ Rev.\ D {\bf 22}, 2227 (1980).

\bibitem{Schechter:1981cv}
  J.~Schechter and J.~W.~F.~Valle,
  Phys.\ Rev.\ D {\bf 25}, 774 (1982).


\bibitem{Ke:2014hxa}
  H.~W.~Ke, T.~Liu and X.~Q.~Li,
  Phys.\ Rev.\ D {\bf 90}, 053009 (2014)  [arXiv:1408.1315 [hep-ph]].  

\bibitem{Liu:2013oxa}
  J.~Y.~Liu and S.~Zhou,
  Phys.\ Rev.\ D {\bf 87}, no. 9, 093010 (2013)
  [arXiv:1304.2334 [hep-ph]].


\bibitem{Girardi:2014wea}
  I.~Girardi, D.~Meloni, T.~Ohlsson, H.~Zhang and S.~Zhou,
   arXiv:1405.6540 [hep-ph].  

\bibitem{Kisslinger:2013sba}
  L.~S.~Kisslinger,
  arXiv:1309.4983 [hep-ph].

\bibitem{Mention:2011rk}
  G.~Mention, M.~Fechner, T.~.Lasserre, T.~.A.~Mueller, D.~Lhuillier, M.~Cribier and A.~Letourneau,
  Phys.\ Rev.\ D {\bf 83}, 073006 (2011)  [arXiv:1101.2755 [hep-ex]].

\bibitem{Giunti:2013aea}
  C.~Giunti, M.~Laveder, Y.~F.~Li and H.~W.~Long,
  Phys.\ Rev.\ D {\bf 88}, 073008 (2013)
  [arXiv:1308.5288 [hep-ph]].

\bibitem{Giunti:2011cp}
  C.~Giunti and M.~Laveder,
  Phys.\ Lett.\ B {\bf 706}, 200 (2011)
  [arXiv:1111.1069 [hep-ph]].
\bibitem{Hamann:2010bk}
  J.~Hamann, S.~Hannestad, G.~G.~Raffelt, I.~Tamborra and Y.~Y.~Y.~Wong,
  Phys.\ Rev.\ Lett.\  {\bf 105}, 181301 (2010)
  [arXiv:1006.5276 [hep-ph]].

\bibitem{Giusarma:2011ex}
  E.~Giusarma, M.~Corsi, M.~Archidiacono, R.~de Putter, A.~Melchiorri, O.~Mena and S.~Pandolfi,
  Phys.\ Rev.\ D {\bf 83}, 115023 (2011)
  [arXiv:1102.4774 [astro-ph.CO]].

\bibitem{Antusch:2014woa}
  S.~Antusch and O.~Fischer,
   JHEP {\bf 1410}, 94 (2014)  [arXiv:1407.6607 [hep-ph]].  

\bibitem{Escrihuela:2015wra}
  F.~J.~Escrihuela, D.~V.~Forero, O.~G.~Miranda, M.~Tortola and J.~W.~F.~Valle,
  arXiv:1503.08879 [hep-ph].  

\bibitem{Babu:2002dz}
  K.~S.~Babu, E.~Ma and J.~W.~F.~Valle,
  Phys.\ Lett.\ B {\bf 552}, 207 (2003)
  [hep-ph/0206292].
\bibitem{Morisi:2013qna}
 D.~V.~Forero, S.~Morisi,  J.~C.~Rom?o and J.~W.~F.~Valle,
  Phys.\ Rev.\ D {\bf 88}, no. 1, 016003 (2013)
  [arXiv:1305.6774 [hep-ph]].
\bibitem{Morisi:2012fg}
  S.~Morisi and J.~W.~F.~Valle,
  Fortsch.\ Phys.\  {\bf 61}, 466 (2013)
  [arXiv:1206.6678 [hep-ph]].




\end{thebibliography}
\end{document}